\documentclass[sn-standardnature]{sn-jnl}

\usepackage{etoolbox}
\usepackage{caption}
\usepackage[flushleft]{threeparttable}
\usepackage{abstract}  
\renewcommand{\abstractname}{}    

\jyear{2023}%

\theoremstyle{thmstyleone}%
\theoremstyle{thmstyletwo}%

\theoremstyle{thmstylethree}%

\begin{document}

\title[TRAPPIST-1 b thermal emission]{Thermal emission from the Earth-sized exoplanet TRAPPIST-1 b using JWST}


\author*[1]{\fnm{Thomas P.} \sur{Greene}}\email{tom.greene@nasa.gov}

\author[1,2]{\fnm{Taylor J.} \sur{Bell}}\email{bell@baeri.org}

\author[3,4]{\fnm{Elsa} \sur{Ducrot}}\email{elsa.ducrot@cea.fr}

\author[3]{\fnm{ Achr{\`e}ne} \sur{Dyrek}}\email{achrene.dyrek@cea.fr}

\author[3]{\fnm{Pierre-Olivier} \sur{Lagage}}\email{pierre-olivier.lagage@cea.fr}

\author[5]{\fnm{Jonathan J.} \sur{Fortney}}\email{jfortney@ucsc.edu}

\affil*[1]{\orgdiv{Space Science and Astrobiology Division}, \orgname{NASA's Ames Research Center}, \orgaddress{\street{M.S. 245-6}, \city{Moffett Field}, \postcode{94035}, \state{CA}, \country{USA}}}

\affil[2]{\orgdiv{Bay Area Environmental Research Institute}, \orgname{NASA's Ames Research Center}, \orgaddress{\street{M.S. 245-6}, \city{Moffett Field}, \postcode{94035}, \state{CA}, \country{USA}}}

\affil[3]{\orgname{Université Paris-Saclay, Université Paris-Cité, CEA, CNRS, AIM}, \orgaddress{\city{Gif-sur-Yvette}, \postcode{91191},  \country{France}}}

\affil[4]{\orgname{Paris Region Fellow, Marie Sklodowska-Curie Action}}

\affil[5]{\orgdiv{Department of Astronomy and Astrophysics}, \orgname{University of California, Santa Cruz}, \orgaddress{\city{Santa Cruz}, \postcode{95064}, \state{CA}, \country{USA}}}


\maketitle
\begin{abstract}
\bf
The TRAPPIST-1 system is remarkable for its seven planets that are similar in size, mass, density, and stellar heating to the rocky planets Venus, Earth, and Mars in our own Solar System \cite{2017Natur.542..456G}. All TRAPPIST-1 planets have been observed with the transmission spectroscopy technique using the Hubble or Spitzer Space Telescopes, but no atmospheric features have been detected or strongly constrained \cite{2018AJ....156..218D, 2018NatAs...2..214D, 2018AJ....156..178Z, 2022A&A...665A..19G}. TRAPPIST-1 b is the closest planet to the system's M dwarf star, and it receives 4 times as much irradiation as Earth receives from the Sun. This relatively large amount of stellar heating suggests that its thermal emission may be measurable. Here we present photometric secondary eclipse observations of the Earth-sized TRAPPIST-1 b exoplanet using the F1500W filter of the MIRI instrument on JWST. We detect the secondary eclipse in each of five separate observations with 8.7-sigma confidence when all data are combined. These measurements are most consistent with re-radiation of the TRAPPIST-1 star’s incident flux from only the dayside hemisphere of the planet. The most straightforward interpretation is that there is little or no planetary atmosphere redistributing radiation from the host star and also no detectable atmospheric absorption from carbon dioxide (CO$_2$) or other species.
\end{abstract}

\vskip 1 cm







The TRAPPIST-1 system has an age of  7.6 $\pm$ 2.2 Gyr \cite{ 2017ApJ...845..110B} and consists of a very cool ($T_{\rm eff}
 = 2566$ K), low-mass star (0.09 solar masses) and seven transiting planets that are 0.75 -- 1.1 Earth radii in size with masses 0.3 -- 1.4 times that of Earth \cite{2017Natur.542..456G, 2021PSJ.....2....1A}. The small size of its star and its nearby location (12 parsecs from Earth) makes its planets amenable to observational characterization studies.

Late M-dwarf stars like TRAPPIST-1 \cite{2006PASP..118..659L} are observed to have frequent X-ray and ultraviolet flares \cite{2017ApJ...851...77R, 2018ApJ...858...55P}. Furthermore, these low-mass stars evolve through an extended high-luminosity pre-main-sequence phase lasting 1 Gyr or more \cite[][]{1998A&A...337..403B}, causing extreme water loss on planets that would later become temperate \cite{2015AsBio..15..119L, 2017MNRAS.464.3728B}. The TRAPPIST-1 planets are also expected to be tidally locked so that their rotations are synchronized to their orbital periods \cite{2017Natur.542..456G}. These phenomena will impact the atmospheres and climates of the TRAPPIST-1 planets, causing them to be significantly different from the terrestrial planets in our Solar System \cite{2007AsBio...7...30T, 2018A&A...612A..86T}. High-energy stellar flares and coronal mass ejections from M dwarfs can also dissociate molecules and drive atmospheric escape \cite[][]{2019AsBio..19...64T, 2020IJAsB..19..136A}. This can be mitigated by planetary magnetic fields if the ejected stellar plasma heats planetary interiors and leads to volcanism and outgassing \cite{2022ApJ...941L...7G}. Tidal synchronization of planetary rotations to orbits will also cause some internal heating that could lead to significant atmospheric outgassing \cite[][]{2022ApJ...933..115K}. Furthermore, tidal synchronization  will cause the TRAPPIST-1 planets to have warm permanent day sides that always face the star and colder night sides that always radiate to free space. 

A number of studies have modeled possible TRAPPIST-1 planetary atmospheres, usually with outgassed secondary compositions that may have undergone further processing via desiccation or reduction. These atmospheres are often dominated by H$_2$O, O$_2$, N$_2$, or CO$_2$ and have higher mean molecular weights than primordial hydrogen-dominated ones \cite{2018A&A...612A..86T, 2018ApJ...867...76L, 2020SSRv..216..100T, 2022ApJ...933..115K}. TRAPPIST-1 planetary transmission spectroscopy observations to date have been precise enough to detect only moderately high concentrations of atomic or molecular constituents in clear hydrogen-dominated atmospheres \cite{2018AJ....156..218D, 2018NatAs...2..214D, 2018AJ....156..178Z, 2022A&A...665A..19G}. These existing observations are insensitive to detecting spectroscopic absorptions in cloudy or high mean-molecular weight atmospheres of planets as small and cool as ones in the TRAPPIST-1 system.

The recently-commissioned JWST should be more sensitive to spectral features and thermal signatures of TRAPPIST-1 planetary atmospheres due to its improved aperture, wavelength range, spectral resolving power, and stability over that of previous observatories \cite{2016MNRAS.461L..92B, 2019AJ....158...27L, 2022ApJ...933..115K}. The inner TRAPPIST-1 planets are warm enough to potentially detect their thermal emission with mid-infrared secondary eclipse observations \cite{2015PASP..127..941C}. TRAPPIST-1 b, the innermost planet, has the highest expected equilibrium temperature T$_{\rm eq}$ = 391--400K \citep{2017Natur.542..456G, 2022ApJ...924..134K}. T$_{\rm eq}$ is computed for absorbing all incident stellar radiation on its dayside and re-radiating it isotropically over 4$\pi$ steradians. Measuring the actual dayside temperature would constrain the existence, composition, and circulation of its atmosphere \cite{2019ApJ...886..140K, 2019ApJ...886..141M}. Spitzer Space Telescope observations did not detect the secondary eclipse of TRAPPIST-1 b but did constrain its dayside temperature to below 611 K (3 $\sigma$ upper limit \citep{ 2020A&A...640A.112D}). 

We now report new JWST observations that detect the TRAPPIST-1 b secondary eclipse, measure the planet's dayside temperature, and constrain the properties of any atmosphere it may have. Our program consisted of  photometric observations of five secondary eclipses of TRAPPIST-1 b using the F1500W filter of the JWST Mid-Infrared Instrument (MIRI) \cite{2015PASP..127..584R}. This filter transmits wavelengths over a half-maximum bandpass of 13.5 -- 16.6 micrometers, nearly ideal for maximizing the signal-to-noise ratio of the secondary eclipse for the planet's expected T $\gtrsim 400$ K dayside temperature. Each observation consisted of 377 integrations of 14 FAST1R groups in full-frame mode, with a duration of 4.36 h. Secondary eclipses were observed on 2022 November 8, 12, 20, 24, and December 3 UTC. The dates and times of the observations were computed from a dynamical model of the TRAPPIST-1 planetary system \cite{2021PSJ.....2....1A} and chosen so that there was no contamination by transits or secondary eclipses of other planets in the system.

\begin{figure}
    \centering
    \includegraphics[width=1.0\textwidth]{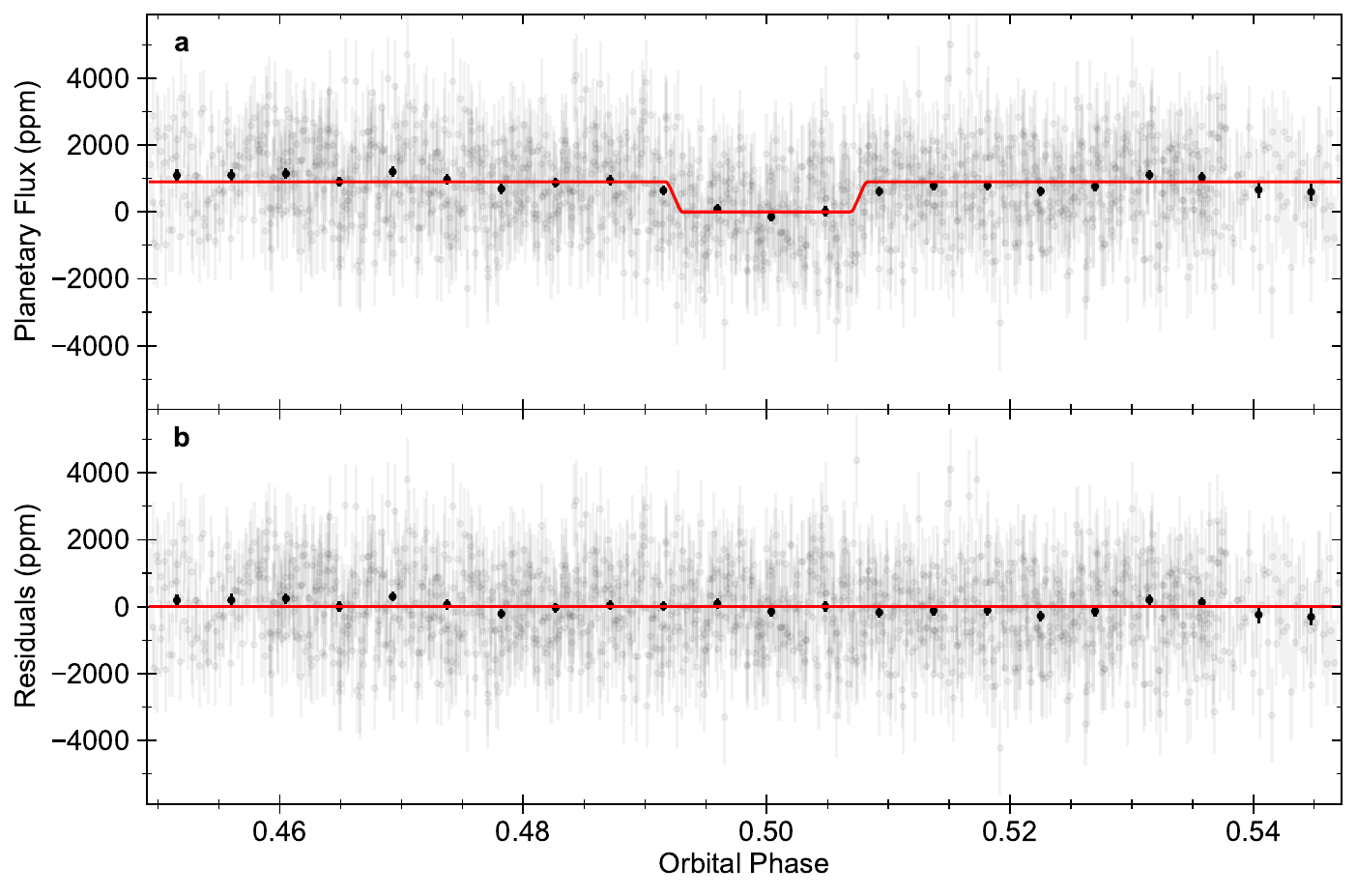}
    \caption{\small Combined TRAPPIST-1 b MIRI F1500W secondary eclipse light curve. 
	{\bf a.} The phase-folded light curves from all of our five observations after the removal of systematic noise based on Joint Fit \#1 (see Methods). Overplotted are data binned to a cadence of 9.7 minutes to more clearly visualize the detection of the eclipse, and the red line shows the fitted astrophysical model. The model fit has eclipse depth $F_{\rm p}/F_{\star} = 861 \pm 99$ ppm.
    {\bf b.} The residuals after fitting the observations. Error bars show 1$\sigma$ uncertainties in both panels.}\label{fig_LC}
\end{figure}

We reduced the data using the \texttt{Eureka!} pipeline which is optimized to minimize noise in JWST time-series data \cite{bell2022}. Secondary eclipse light curves were generated and fit from each observation, and the light curves from all observations were also jointly fit (Methods). The secondary eclipse was detected in each observation, and Figure \ref{fig_LC} shows the combined light curve and fit for all observations.

Next, we convert the measured eclipse depth, $F_{\rm p}/F_{\star} = 861 \pm 99$ ppm, to the measured dayside planetary flux ($F_{\rm p}$) and to its brightness temperature within the JWST MIRI F1500W photometric bandpass. We used calibrated JWST data products to compute a mean stellar flux of $2.589 \pm 0.078$ mJy (Methods). Therefore, TRAPPIST-1 b has a dayside flux  $F_{\rm p} = 2.229 \pm 0.263$ $\mu$Jy. This corresponds to a blackbody brightness temperature $T_{\rm B} = 503\substack{+26 \\ -27}$\, K, computed from Planck's law.
We calculated this $T_{\rm B}$ value for the F1500W filter's effective wavelength of $\lambda = 14.79$ $\mu$m. This wavelength was computed by weighting JWST's wavelength-dependent throughput in this filter by the flux of a PHOENIX stellar model \cite{2013A&A...553A...6H} interpolated to $T_{\rm eff} = 2566$ K to represent the stellar spectrum. 

The apparent observed dayside temperature of a planet in a non-eccentric orbit with zero heat redistribution to its night side is $T_{\rm d} = [2/3(1-A)]^{1/4} T_0$ where $T_0 = T_{\rm eff}(R_\star/a)^{1/2}$ for semi-major axis $a$, bond albedo $A$, stellar effective temperature $T_{\rm eff}$, and stellar radius $R_\star$ \cite{2011ApJ...726...82C}. The planet's substellar point has temperature $T_0$, and all other locations have lower temperatures and contribute less relative area to the observed planet flux in a secondary eclipse observation. For the adopted parameters of TRAPPIST-1 b \cite{2021PSJ.....2....1A} and $A = 0$, we compute $T_{\rm d} = 508$\,K, very close to our measured \mbox{$T_{\rm B} = 503\substack{+26 \\ -27}$\,K}.

Any atmospheric circulation that redistributes stellar heating or absorption by atmospheric constituents within the F1500W filter's spectral bandpass will impact the planet's observed secondary eclipse depth and brightness temperature. Model TRAPPIST-1 b planetary surfaces composed of anorhite, basalt, enstatie, feldspar olivine, pyroxene, quartz, or saponite minerals with no gaseous atmospheres are all expected to have secondary eclipse depths and brightness temperatures similar to blackbodies in the MIRI F1500W bandpass \cite{2019AJ....158...27L}. These and other materials could potentially be identified with low-resolution spectroscopic observations at other wavelengths. Furthermore, atmospheres dominated by CO$_2$ or O$_2$ with some CO$_2$ (outgassed or desiccated) with surface pressures as low as 10 bar should have significant absorptions in the F1500W filter bandpass, reducing the observed brightness temperature to 300 K or less when accompanied by efficient heat redistribution over the planet \cite{2019AJ....158...27L, 2018ApJ...867...76L}. Table \ref{table_TB} shows that such low brightness temperatures are inconsistent with our observations. Furthermore, our measured secondary eclipse depth is also inconsistent with the $T_{\rm eq} = 400$ K expected for uniform stellar heat redistribution and isotropic re-radiation even with no absorbing atmospheric species within the spectral bandpass of our observations. 

The heat redistribution of TRAPPIST-1 b atmospheres have been modeled in terms of their surface pressures P and effective gray optical depths $\tau$ \cite{2022ApJ...924..134K}. Table \ref{table_TB} shows that our observations are consistent with a thin atmosphere with little heat distribution; the flux for the surface pressure P = 0.1 bar and gray optical depth $\tau = 0.1$ model \cite{2022ApJ...924..134K} is well within 1 $\sigma$ of our measurement. However, the atmosphere model with 10 times these values recirculates too much heat and has a dayside temperature that is too low to be clearly consistent with our observations. 

A coupled atmosphere-interior model that includes heating from tidal effects, interior processes, and instellation predicts isotropic surface temperatures for different TRAPPIST-1 b scenarios at its present age \cite[][]{2022ApJ...933..115K}. The median isothermal surface temperature of the 50 hottest and lowest albedo PACMAN models with negligible final atmospheres (P $<$ 0.1 bar) was 446 K \cite[][]{2022ApJ...933..115K} (and J. Krissansen-Totton private communication). We converted this mean isotropic surface temperature to an apparent observed dayside temperature T$_{\rm day}$ = 534 K using the geometric and thermal conventions for zero redistribution of stellar heating to the night side as was done in defining $T_{\rm d}$ above \cite{2011ApJ...726...82C}. This should be regarded as the maximum T$_{\rm day}$ possible from this particular internal heating model given that it is derived from only the hottest 10\% of model runs that resulted in negligible atmospheres. Our observations are consistent with this prediction (see Table \ref{table_TB}). 

\begin{table}[h]
\begin{threeparttable}
\caption{\bf TRAPPIST-1 b Observations and Model Comparison\label{table_TB}}%
\begin{tabular}{lrrrr}
\hline
Atmospheric Model & Ref. & P$_{\rm surf}$(bar)& Predicted $T_{\rm B}$\footnotemark[1] (K) & Difference ($\sigma$)\footnotemark[2] \\ [0.5ex]
\hline
96.5\% CO$_2$ & \cite{2019AJ....158...27L} &  10, 92 & 290 &  -6.7 \\
95\% O$_2$ + 0.5 bar CO$_2$ & \cite{2019AJ....158...27L} & 10, 100 & 303 & -6.4 \\
Isotropic T$_{\rm eq}$ & \cite{2017Natur.542..456G} & N/A & 400 & -3.6 \\
Gray $\tau = 1.0$ & \cite{2022ApJ...924..134K} & 1 & 445 & -2.1 \\
Gray $\tau = 0.1$ & \cite{2022ApJ...924..134K} & 0.1 & 495 & -0.3 \\
0 redistribution & \cite{2011ApJ...726...82C} & 0 & 508 & 0.2 \\
0 redistrib. + internal & \cite{2022ApJ...933..115K} & $<0.1$ & 534 & 1.2 \\
\hline
\end{tabular}
\begin{tablenotes}
\footnotesize
\item 1. $T_{\rm B}$ in MIRI F1500W filter, when applicable.
\item 2. Difference between predicted and observed $\lambda = 15$ $\mu$m flux in units of measured $\sigma$. 
\end{tablenotes}
\end{threeparttable}
\end{table}

\begin{figure}%
\centering\includegraphics[width=1.0\textwidth]{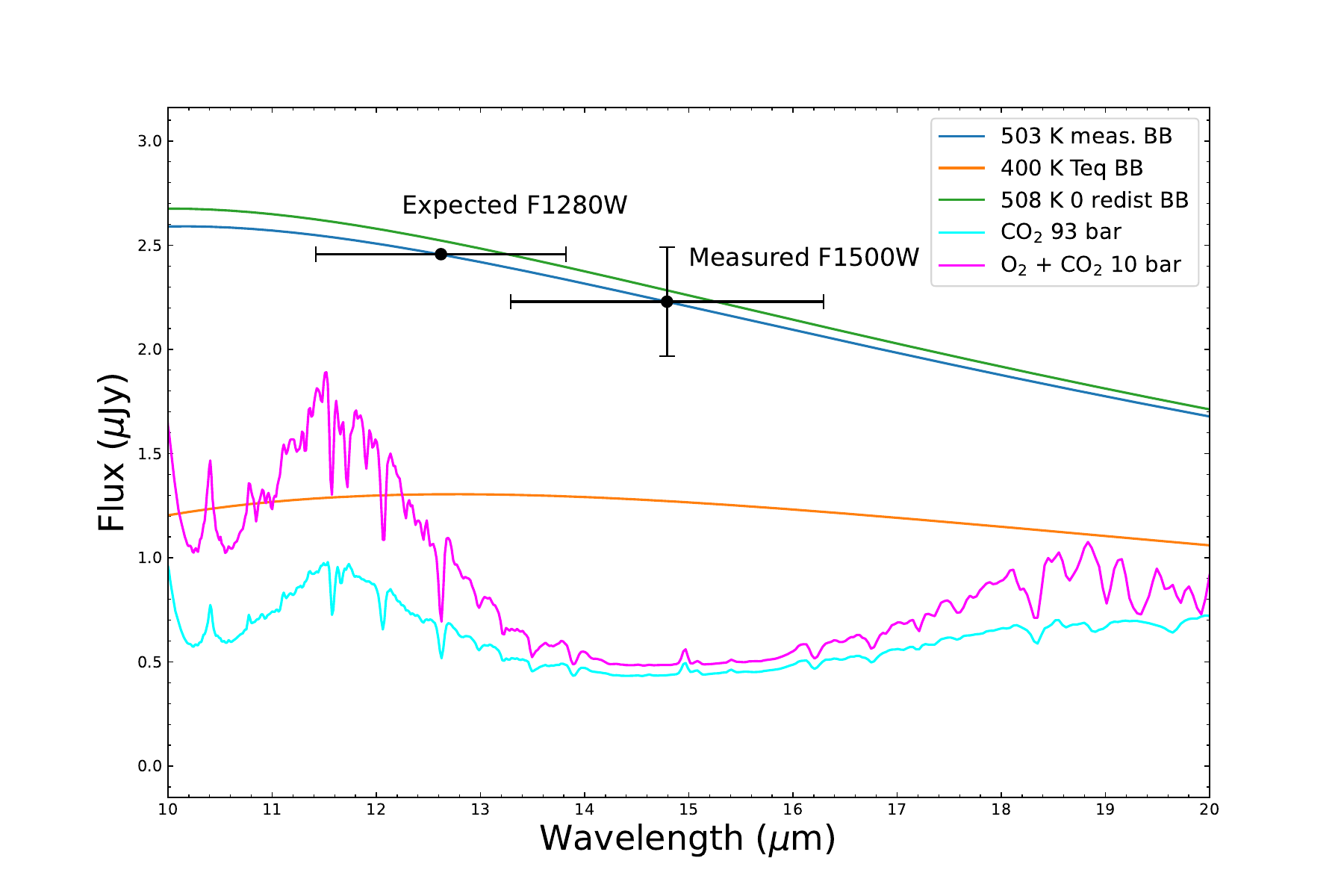}
\caption{\small TRAPPIST-1 b F1500W measured flux and spectral models. The blackbody curves represent the measured T$_{\rm B}$ = 503 K dayside temperature, the 508 K apparent dayside temperature predicted for zero heat redistribution and no internal heating, and T$_{\rm eq}$ = 400 K temperature for isotropic redistribution of stellar heating. The flux expected in the upcoming observations in the MIRI F1280W filter is also shown assuming the planet emits like a T$_{\rm B}$ = 503 K blackbody. The widths of the F1280W and F1500W markers represent their transmission half-amplitude bandpasses, and the vertical error bar of the F1500W point represents its 1$\sigma$ uncertainty. Emergent spectra from 93 bar CO$_2$ and 10 bar outgassed O$_2$ with 0.5 bar CO$_2$ atmospheres are also plotted \cite{2018ApJ...867...76L, VPLmodels}. 
}\label{fig_model_spec}
\end{figure}

Our observations are most consistent with near-zero bond albedo and little-to-no heat redistribution from the dayside to the nightside of TRAPPIST-1 b. The most likely explanation for this finding is that the planet absorbs nearly all of the incident stellar flux and does not have a high-pressure or optically-thick atmosphere. Figure \ref{fig_model_spec} shows the measured TRAPPIST-1 b flux in the F1500W filter along with model spectra for blackbodies, CO$_2$, and O$_2$ + CO$_2$ atmospheres \cite{2018ApJ...867...76L, VPLmodels}. Our data are clearly incompatible with the T$_{\rm eq}$ = 400K, CO$_2$, or O$_2$ + CO$_2$ models. Disfavoring a substantial, high mean molecular weight secondary atmosphere suggests that the null results from the previous transmission spectroscopy observations \cite[][]{2018AJ....156..218D, 2018NatAs...2..214D, 2018AJ....156..178Z} are likely caused by little or no atmosphere of any kind. 
The absence of a substantial atmosphere is also consistent with a coupled atmosphere-interior model that predicts complete atmospheric erosion for TRAPPIST-1 b over its lifetime in approximately half of the model cases run \cite{2022ApJ...933..115K}.

Future observations have the potential to further constrain the heat redistribution and possible atmospheres of TRAPPIST-1 b. Spitzer 3.6 $\mu$m and 4.5 $\mu$m band fluxes were once interpreted as possibly due to CO$_2$ in emission at the 2$\sigma$ level \cite{2018MNRAS.475.3577D}, but this was not seen in additional observations \cite{2020A&A...640A.112D}. JWST secondary eclipse measurements in the MIRI F1280W filter (JWST program 1279) may better constrain the dayside brightness temperature and determine how well the planet's emission does or does not follow a blackbody emission or atmospheric absorption spectrum (see Figure \ref{fig_model_spec}).  Thermal emission phase curve observations that capture most of the planet's orbit (P = 36 hours) could also measure the contrast between its day and night sides. This would allow measuring heat transfer and could further determine whether there is a tenuous atmosphere, as was done with the Spitzer Space Telescope for the much hotter rocky planet LHS 3844 b \cite{2019Natur.573...87K}. Secondary eclipse or phase curve thermal emission observations of additional similar planets including others in the TRAPPIST-1 system (e.g., JWST program 2304) could also reveal systematic properties of M dwarf terrestrial planets and how they differ from Solar System ones.


\pagebreak

\section*{Methods}\label{sec11}

\subsection*{Data and analysis overview}

These JWST MIRI F1500W time-series observations feature significant background signals, dominated by thermal emission from the JWST primary mirror and sunshade within the F1500W filter bandpass \cite{2012SPIE.8442E..3BL, 2022arXiv221109890R}. At the start of each of the five exposures, we measured a mean background of 19,837--22,779 electrons per pixel in an annulus of radius 12--30 pixels per 38.9\,s integration. These values assume a gain of 3.1 electrons per digital number (hereafter DN). This background level corresponds to approximately 45\% of the total flux within a five-pixel radius photometric aperture centered on the TRAPPIST-1 star.
Therefore, we note that it is important to subtract the background flux accurately to maximize the quality of these and perhaps other long-wavelength JWST data.

We now describe how we analyzed the raw image data to measure the secondary eclipse parameters of TRAPPIST-1 b in two different ways. The descriptions below refer to calibration pipelines and other software whose code and citations appear in the ‘Code availability’ section.


\subsection*{\texttt{Eureka!} Analyses: Images to light curves}

We reduced our observations using version 0.9 of the \texttt{Eureka!} data analysis pipeline \cite{bell2022}, with \texttt{jwst} version 1.8.3 \cite{JWSTpipe}, CRDS version 11.16.16 \cite{crds}, and using the ``jwst\textunderscore 1019.pmap" CRDS context. We started from the \textunderscore uncal.fits files and ran Stages 1-4 of \texttt{Eureka!}. Stage 1 was first run using the default \texttt{jwst} settings with the exception of increasing the jump step's cosmic ray rejection threshold to 10 sigma to avoid an excessively large number of false positive cosmic ray detections. The only change to the Stage 2 defaults were in turning off the photom step except when computing the calibrated stellar flux; turning off the photom step makes it easier to understand how close our reduction is to the photon noise limit. 

In \texttt{Eureka!}'s Stage 3, we performed aperture photometry using source radii ranging from 4--12 pixels. In some of our observations, we found that using a large aperture (e.g. a 9 pixel radius which encircles the first Airy ring) resulted in a smaller eclipse depth as well as noisier residuals; we attributed this to imperfect background subtraction across the large source aperture resulting in a dilution of our eclipse depth. Ultimately, we found that an aperture radius of 5 pixels (which encircles the Airy disk) offered the lowest scatter in our residuals by maximizing the encircled starlight and minimizing the encircled background flux. We also considered a range of annuli to determine the background in each frame which was important as the background flux varies significantly over the duration of the observations. We found that the choice of background annulus had little impact on the resulting eclipse depth, but an annulus that started near the end of the first Airy ring and included many background pixels helped to reduce the scatter in our residuals; ultimately we chose an annulus spanning 12--30 pixels from the centroid. We also masked the values marked as DO NOT USE in the DQ array, 5$\sigma$ clipped background flux outliers and linearly interpolated bad pixels. We computed the centroid for each integration using a 2D Gaussian fit with an initial value for the centroid of x=698, y=516 and using only the 20x20 grid of pixels surrounding this first value. In \texttt{Eureka!}'s Stage 4, we sigma clipped 10$\sigma$ outliers compared to the local median flux computed using a boxcar filter with a width of 10 integrations. 

We also performed an alternative data reduction that was identical to the one described thus far except we set the \texttt{ramp\_fitting\_weighting} to \texttt{uniform} in Stage 1. All groups in each integration of the data are limited by photon noise (background and star signal) and not read noise, so it is logical to weight them uniformly instead of using the default \texttt{optimal} value which is optimized for faint sources and low backgrounds.

\subsection*{Calibrated Stellar Flux}
We measured the stellar flux of TRAPPIST-1 from  \texttt{\_calints.fits} data products that we downloaded from the JWST archive. These had undergone standard processing with calibration version 1.8.2 and data processing version 2022\_4a. After discarding the first 40 integrations of each observation, we computed the flux in a 15 pixel radius aperture and subtracted the background in a 16--35 pixel annulus centered on the star. We applied corrections for starlight outside of the measurement aperture and inside the background annulus ($\sim$4\% each; \cite{2022arXiv220705632R, jdox_miri_apcorr}) and applied the pixel scale of $2.844\times10^{-13}$~sr/pixel. The calibrated flux varied by $<0.4$\% ($\sigma = 0.004$ mJy) between the five observations, and there is an additional absolute calibration uncertainty $\sim$3\% (0.078 mJy; K. Gordon / Space Telescope Science Institute private communication). We add the systematic and observational uncertainties in quadrature to get an observed stellar flux of $2.589 \pm 0.078$ mJy. This is 13\% lower than than the flux expected from a Phoenix / BT-Settl model produced by the  \texttt{pysynphot} package for the stellar parameters and normalized to the star's observed 2MASS J = 11.35 mag flux.

\subsection*{Secondary eclipse depths}
To begin, we first individually fit each of our visits to assess the data quality of each visit and assess the evidence for astrophysical variability. We performed two independent fits to the observations to ensure the reproducibility of our conclusions. Individual and joint fit \#1 used the standard-processed Stage 1--4 data, while individual and joint fits \#2 used the alternatively processed data with \texttt{ramp\_fitting\_weighting=uniform}. We describe the method used in each fit below.

\paragraph{Individual Fits \#1}
For the first fit to the individual observations, we used \texttt{Eureka!} to fit a uniform-surface eclipse depth model using \texttt{starry} \cite{starry_v1.2.0}, a linear trend in time, and we linearly decorrelated against the star's x and y position as well as the star's x and y PSF width as determined using the Gaussian centroiding method in \texttt{Eureka!}'s Stage 3. We also inflated the estimated white noise level during the fit to account for noise above the photon limit. The first $\sim$100 integrations showed a moderate ramp shape that was likely caused by detector settling, and we decided to remove these data rather than try to fit them with an exponential decay model. To fit the observations, we used PyMC3's No U-Turns Sampler \cite{pymc3} and ran two chains with a target acceptance rate of 0.95 with each chain taking 5000 tuning steps and then 2000 posterior samples. The Gelman-Rubin statistic \cite{GelmanRubin1992} was used to ensure the chain had converged. For astrophysical priors, we took $R_{\rm p}/R_\star = 0.08590$, $R_\star = 0.1192\,R_{\odot}$ (to account for light travel time), $F_{\rm p}/F_\star = 0 \pm 0.01$, $P = 1.5108794$ days, $t_0 = 59890.5150313 \pm 0.005$ (BMJD\textunderscore TDB), $i = 89.728^{\circ}$, $a/R_\star = 20.843$, and $e = 0$ \cite{2021PSJ.....2....1A}. From these independent fits, only the eclipse from November 20th showed evidence for a small amount of unmodelled red noise in its Allan variance plot \citep{Allan1966}; this unmodelled red noise is likely a slight downward ramp at the start of this observation (see Extended Data Figure \ref{fig:individual_lcs}).

Note that our $t_0$ prior comes from numerical predictions which account for planet--planet interactions which significantly perturb the planets' orbits from static, circular orbits \cite{2021PSJ.....2....1A}. To ensure that any inaccuracies in those predictive models do not bias our eclipse depth, we inflated the uncertainty on $t_0$ compared to the published uncertainty \cite{2021PSJ.....2....1A} to search through a larger region of parameter space. Reassuringly, all of our individual eclipse timings were within error of each other and within 2 $\sigma$ of the predicted timings \cite{2021PSJ.....2....1A} assuming zero eccentricity. Each observation constrained the eclipse timing to within $\sim$3 minutes, or $\sim$8\% of the eclipse duration. The mid-eclipse times for the five observations are given in Extended Data Table 1. 

Our five independently fit observations gave eclipse depths of $790\pm210$ ppm, $510\pm210$ ppm, $950\pm170$ ppm, $820\pm220$ ppm, and $829\pm200$ ppm. Each of these observations is consistent with an average eclipse depth of 795 ppm, and we find no evidence for variability in the eclipse depth of TRAPPIST-1b.

\paragraph{Individual Fits \#2}
For the second fit to the individual observations, we used the \texttt{Fortran} code \texttt{trafit} which is an updated version of the adaptive Markov-Chain Monte Carlo (MCMC) code described in \cite{Gillon2010,Gillon2012,Gillon2014}. It uses the eclipse model of \cite{Mandel2002} as a photometric time-series, multiplied by a baseline model to represent the other astrophysical and instrumental systematics that could produce photometric variations. 
First, we fit all visits individually. We tested a large range of baseline models to account for different types of external sources of flux variations/modulations (instrumental and stellar effects). This includes polynomials of variable orders in: time, background, PSF position on the detector (x,y) and PSF width (in x and y). Once the baseline was chosen, we ran a preliminary analysis with one Markov chain of 50,000 steps to evaluate the need for re-scaling the photometric errors through the consideration of a potential under- or over-estimation of the white noise of each measurement and the presence of time-correlated (red) noise in the light curve. After re-scaling the photometric errors we ran two Markov chains of 100,000 steps each to sample the probability density functions of the parameters of the model and the system's physical parameters, and assessed the convergence of the MCMC analysis with the Gelman \& Rubin statistical test \citep{GelmanRubin1992}.
Our jump parameters were the following :
 the mass of the star $M_{\star}$ with a normal prior distribution $\mathcal{N}(0.0980,0.0023^{2})M_{\odot}$;
the radius $R_{\star}$ with a normal prior distribution $\mathcal{N}(0.1192,0.0033^{2})R_{\odot}$;
the effective temperature $T_{eff}$ with a normal prior distribution $\mathcal{N}(2520,39^{2})K$;
and the metallicity [Fe/H] of the star with a normal prior distribution $\mathcal{N}(0.0535,0.0880^{2})$;
 the impact parameter of the planet $b$ with a normal prior distribution $\mathcal{N}(0.25,0.11^{2})$;
 and $T_0$ with normal priors and inflated errors similarly to Fit \#1; 
 and $F_{\rm p}/F_{\star}$ with a broad uniform prior. 
All normal priors were taken from \cite{ 2020A&A...640A.112D} except for the transit timing that were derived from the dynamical model predictions by \cite{2021PSJ.....2....1A}. All of our individual eclipse timings were also within 2 $\sigma$ of this dynamical model's predicted timings assuming zero eccentricity. 

For the five individual fits the Gelman-Rubin statistic was lower than 1.003 for every jump parameter measured across the two chains, indicating good convergence. The eclipse depths that we measured for visit 1 to 5 are: $994^{-193}_{+187}$ppm, $691^{-169}_{+166}$ppm, $798^{-282}_{+284}$ppm, $821^{-229}_{+243}$ppm, $736^{-259}_{+266}$ppm.

\subsubsection*{Combining the multiple eclipse measurements}
Finally, we simultaneously fit all five of the eclipse observations to allow for better constraints on the timing of eclipse; this is important as any error in the mid-eclipse time will result in a bias toward smaller eclipse depths as some of the eclipse signal will included in the baseline and vice versa.

\paragraph{Joint Fit \#1}
For the first joint fit to all observations, we followed the same method as the Individual Fits \#1 with the exception of fitting the period using a Gaussian prior of $P = 1.5108794\pm0.000006$ \cite{2021PSJ.....2....1A} and assuming a shared $F_{\rm p}/F_\star$, $P$, and $t_0$ between all visits; all other fitted variables were allowed to vary freely between the different visits. We again fit the observations using PyMC3's No U-Turns Sampler \cite{pymc3} and ran two chains with a target acceptance rate of 0.90 with each chain taking 6000 tuning steps and then 3000 posterior samples. Again, the Gelman-Rubin statistic \cite{GelmanRubin1992} was used to ensure the chain had converged. From this joint fit, again only the eclipse from November 20th showed evidence for a small amount of unmodelled red noise in its Allan variance plot \citep{Allan1966}. This shared fit constrained the eclipse time to within 36 seconds ($t_0 = 2459891.01487704778 \pm 0.00042$; or $\sim$1\% of the eclipse duration) and is in close agreement with the numerical predictions \cite{2021PSJ.....2....1A}. Our fitted orbital period ($P = 1.5108793 \pm 0.0000062$ days) also remained very consistent with the prior value \cite{2021PSJ.....2....1A}. Overall, we find that the numerical, planet--planet interaction model \cite{2021PSJ.....2....1A} based on observations taken three or more years before our JWST observations was able to predict the timing of TRAPPIST-1b's eclipses with excellent accuracy and precision. We also find that the eclipse duration is consistent with the planet's published transit duration, but our limited data allow constraining this time to only $\sim$10\%.

The individual raw and de-trended light curves as well as the joint model fits to our phase-folded observations are shown in Extended Data Figure \ref{fig:individual_lcs}. With this joint fit, we measure an average eclipse depth of $899 \pm 91$ ppm. While this average eclipse depth is significantly higher than that from the individual fits, this seems to be the result of a bias towards smaller eclipse depths in the individual fits as a result of higher uncertainty on the eclipse timing. Indeed, allowing the eclipse depths to vary between visits while only requiring the time of eclipse to remain consistent between visits gives an average eclipse depth entirely consistent with our joint fit. The phase-folded light curves from all observations after systematic noise removal are shown with the joint eclipse model fit in Fig. \ref{fig_LC} (Main).

\paragraph{Joint Fit \#2}
For the second joint fit to all observations, we used \texttt{trafit} with the baseline models derived from our individual fit for each light curve (see Individual Fits \#2). Again we did a preliminary run of one chain of 50,000 steps to estimate the correction factors that we then apply to the photometric error bars and then a second run with two chains of 100,000 steps. The jump parameters were the same as for the individual fit \#2 expect from the fact that we fixed $T_0$ and allowed for transit timing variations to happen for each visit (each transit TTV has an unconstrained uniform prior). We used the Gelman-Rubin statistic to asses the convergence of the fit. We measure an eclipse depth of $823^{-87}_{+88}$ ppm from this joint fit.

\subsubsection*{Adopted Eclipse Depth and Noise}
We compute a single mean secondary eclipse depth $F_{\rm p}/F_{\star} = 861$ ppm from the two separate jointly-fit values of 899 and 823 ppm. We adopt the uncertainty 99 ppm = $(\sigma_1^2 + \sigma_2^2)^{1/2}$ where $\sigma_1$ = 91 ppm is the larger of the two jointly-fit random uncertainties, and $\sigma_2$ = 38 ppm is an estimate of the systematic error equal to the difference between either jointly-fit depth and their mean. The random noise value of 91 ppm is $\sim$1.3 times the photon noise we measure from the data and also estimate from an in-flight JWST MIRI performance model. This excess noise factor is similar to one measured for the MIRI low resolution spectroscopy (LRS) time-series observing mode which uses the same detector and signal-chain electronics \cite{2022arXiv221116123B}.


\subsubsection*{Data Availability}
The data used in this paper are associated with JWST GTO program 1177 (observations 7 -- 11) and will be publicly available from the Mikulski Archive for Space Telescopes (https://mast.stsci.edu) at the end of their one-year exclusive access periods. \\

\subsubsection*{Code Availability}
We used the following codes to process, extract, reduce and analyse the data: STScI JWST Calibration pipeline \cite{JWSTpipe}, Eureka! \cite{bell2022}, Emcee \cite{2013PASP..125..306F}, starry \cite{starry_v1.2.0}, PyMC3 \cite{pymc3}, PySynphot \cite{pysynphot}, and the standard Python libraries numpy \cite{harris2020array}, astropy \cite{astropy:2022}, and matplotlib \cite{Hunter:2007}. These were incorporated into custom python notebooks for data analysis. These notebooks are available on request. The notebooks were developed by a NASA employee and are not available publicly until approved by NASA.\\

\makeatletter
\apptocmd{\thebibliography}{\global\c@NAT@ctr 35\relax}{}{}
\makeatother

\vskip 1 cm
\backmatter

\bmhead{Acknowledgments}
We thank E. Schlawin, M. Gillon, V. Parmentier, and E. Rauscher for discussions and two anonymous referees for comments that helped improve the manuscript. This work is based on observations made with the NASA/ESA/CSA James Webb Space Telescope. The data were obtained from the Mikulski Archive for Space Telescopes at the Space Telescope Science Institute, which is operated by the Association of Universities for Research in Astronomy, Inc., under NASA contract NAS 5-03127 for JWST. These observations are associated with program JWST-GTO-1177. We thank the MIRI instrument team and the many other people who contributed to the fantastic success of JWST. T.P.G and T.J.B. acknowledge funding support from the NASA Next Generation Space Telescope Flight Investigations program (now JWST) via WBS 411672.07.04.01.02. 
This material is based upon work supported by NASA'S Interdisciplinary Consortia for Astrobiology Research (NNH19ZDA001N-ICAR) under award number 19-ICAR19\_2-0041 (for J.J.F) and NASA WBS 811073.02.12.04.71 (for T.P.G.).
P.O.L. acknowledges funding support from CNES. E.D. acknowledges support from the innovation and research Horizon 2020 program in the context of the Marie Sklodowska-Curie subvention 945298.








\bmhead{Author contributions}
T.P.G. provided program leadership, devised the observational program, led setting the observation parameters, contributed to the data analysis, led the interpretation of results, and led the writing of the manuscript. 
T.J.B. verified the observing parameters and led the data analysis.
E.D. checked and contributed to the observing parameters, contributed to the data analysis, and contributed to the interpretation of the results.
A. D. contributed to the data reduction.
P.-O.L. contributed to the design of the observational program, contributed to setting the observing parameters, commented on the draft manuscript, and contributed to the data analysis.
J.J.F. contributed to the design of the observational program, provided information for the manuscript, and helped interpret the results.\\

\flushleft The authors declare no competing interests.\\

{ 
\flushleft  Correspondence and requests for materials should be addressed to Thomas Greene.\\
}
 

\noindent

\pagebreak
\begin{appendices}

\section*{Extended Data}\label{Extended Data}



\begin{figure}[h!]
    \captionsetup{name=Extended Data Figure}
    \centering
    \includegraphics[width=\textwidth]{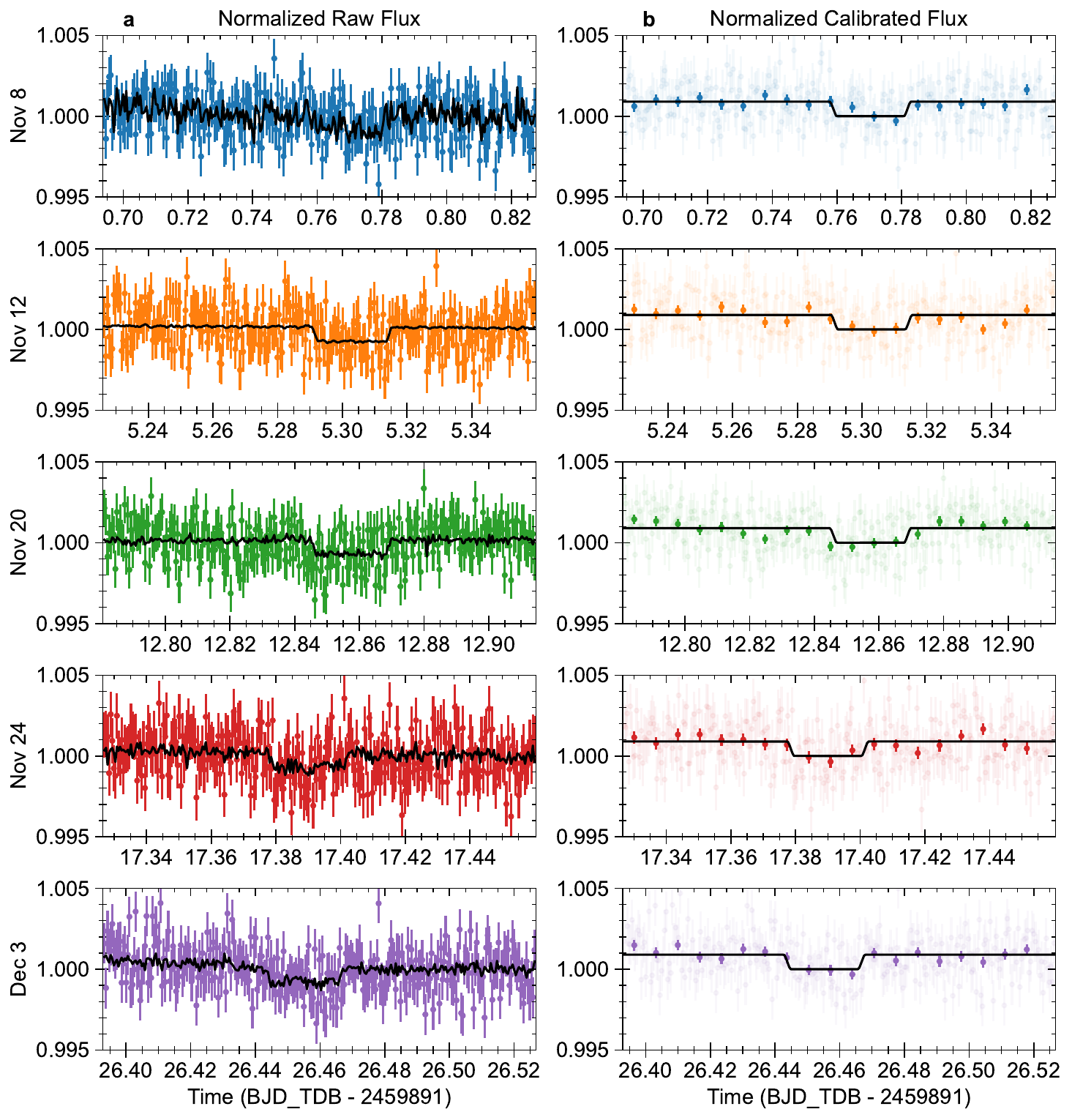}
    \caption{\small Light curves of individual observations and the Joint Fit \#1 models. {\bf a.} The raw light curves from each of our five visits normalized by their median value are shown in color, while the fitted model from Joint Fit \#1 including systematic noise is shown using a black line. The date in UT of each visit is indicated by the y-axis labels. Each visit did not start at the same orbital phase, so the apparent movement in the eclipse time is simply caused by when the observations began.
    {\bf b.} The same data and model from each visit after the removal of systematic noise. Overplotted are data binned at a cadence of 9.7 minutes (14 integrations) to more clearly visualize the detection of the eclipse in each visit. All error bars show 1$\sigma$ uncertainties in both panels.}
    \label{fig:individual_lcs}
\end{figure}

\begin{table}[h]
\captionsetup{name=Extended Data Table}
\begin{center}
\begin{threeparttable}
\caption{\bf Mid-eclipse Times\label{times-table}}%
\renewcommand{\arraystretch}{1.4}
\begin{tabular}{crl}
\hline
Observation & UT date (2022) & mid-eclipse time\footnotemark[1] (BJD TDB) \\ [0.5ex]
\hline
1 &  8 November & $2459891.7711\substack{+0.0017 \\ -0.0010}$ \\
2 & 12 November & $2459896.3038\substack{+0.0018 \\ -0.0011}$ \\ 
3 & 20 November & $2459903.8559\substack{+0.0011 \\ -0.0015}$ \\ 
4 & 24 November & $2459908.3893\substack{+0.0012 \\ -0.0015}$ \\
5 & 3 December  & $2459917.4563\substack{+0.0009 \\ -0.0010}$\, \\
\hline
\end{tabular}
\begin{tablenotes}
\small
\item 1. Stated uncertainties are 1 $\sigma$.
\end{tablenotes}
\end{threeparttable}
\end{center}
\end{table}



\end{appendices}




\end{document}